\documentclass[prl,twocolumn,reprint,preprintnumbers,
nofootinbib,superscriptaddress,longbibliography,nobibnotes]{revtex4-2}

\usepackage{amsmath,amssymb,bm}
\usepackage{slashed}
\usepackage{graphicx}
\usepackage{xcolor}
\usepackage{hyperref}
\usepackage{microtype}

\hypersetup{colorlinks=true,linkcolor=blue,citecolor=blue,urlcolor=blue}

\newcommand{\gev}{\,{\rm GeV}}
\newcommand{\mev}{\,{\rm MeV}}
\newcommand{\kev}{\,{\rm keV}}
\newcommand{\tev}{\,{\rm TeV}}
\newcommand{\cm}{\,{\rm cm}}
\newcommand{\fd}{F_D}
\newcommand{\AD}{A_D}
\newcommand{\qroi}{q_{\rm ROI}}

\begin{document}

\title{Dark Diffraction at LZ from a Screened Neutral Composite Baryon}

\author{Hyunjoo Jung}
\email{hj.jung00@yonsei.ac.kr}
\affiliation{Department of Physics, Yonsei University, Seoul 03722, South Korea}

\author{Seong Chan Park}
\email{sc.park@yonsei.ac.kr}
\affiliation{Department of Physics, Yonsei University, Seoul 03722, South Korea}
\affiliation{School of Physics, Korea Institute for Advanced Study, 85 Hoegi-ro, Seoul 02455, South Korea}

\date{\today}

\begin{abstract}
The LUX-ZEPLIN (LZ) experiment recently reported a nuclear-recoil-like candidate at $E_R=248\pm23_{\rm stat}\pm23_{\rm sys}\kev$. We show that such an energetic recoil can arise from the \emph{elastic} form factor of composite dark matter, without invoking an inelastic mass splitting or a boosted population. A neutral baryon of $N$ dark quarks can have a charge form factor that is suppressed at low momentum transfer but recovers at higher momentum transfer, providing a momentum-space band-pass filter. Modeling the form factor with one Gaussian per charged constituent species, we prove that two terms cannot produce a hard recoil spectrum and find that three yield too few events in the LZ signal window, while four suffice. A confining $SU(4)_D\times U(1)_D$ gauge theory provides a concrete realization through a neutral bosonic baryon made of four dark quarks. For a $200\gev$ benchmark normalized to one accepted event, we predict approximately $1.5$ true recoils beyond the current LZ region of interest. The form factor also produces diffraction nodes at fixed momentum transfers, placing them at recoil energies that scale as $1/m_A$ across target nuclei.
\end{abstract}

\maketitle

The LZ Collaboration has extended its nuclear-recoil search to approximately
$270\kev$ and reported one nuclear-recoil-like candidate at
$E_R=248\pm23_{\rm stat}\pm23_{\rm sys}\kev$, with a maximum local
significance of $3.4\sigma$ among the models LZ tested ($2.6\sigma$
global)~\cite{LZ2026}.  The momentum transfer,
$q_\star\simeq\sqrt{2m_{\rm Xe}E_R}\simeq246\mev$, is far above where halo
dark matter ordinarily scatters.  Interpretations so far invoke inelastic
(endothermic) dark matter in many
guises~\cite{SuYangYang2026,DiMauro2026,WangXiao2026,Yamashita2026,
LeeRandall2026,Yuan2026}, most predictively a $Z$-mediated
Higgsino~\cite{FreeseTheodosopoulos2026,Yin2026,Bisal2026,Cheung2026} now
confronted by the empty high-energy sideband~\cite{Rodd2026} and by solar
capture~\cite{DiMauroShaikh2026,Langhoff2026}, or exothermic, boosted,
absorbed, or neutrino-up-scattered dark matter and nontrivial xenon
responses~\cite{BaerBarger2026,Kannike2026,Alhazmi2026,Heikinheimo2026,
LouLu2026,JeesunMajumdar2026,Khan2026,ParkBoosted2020,ParkNuBDM2021,
ParkMeV2008,ParkLightVector2021}; elastic neutrino scattering cannot account
for the event~\cite{Chattaraj2026}.

Composite dark matter offers a qualitatively different, elastic possibility.
Internal form factors can strongly distort direct-detection
spectra~\cite{Feldstein2009,Hardy2015}, strongly coupled sectors admit
low-energy descriptions in terms of flavor-resolved
currents~\cite{ParkComposite2023,ParkComposite2026}, and confining $SU(4)$
theories possess stable bosonic baryons studied on the lattice as dark-matter
candidates~\cite{Appelquist2014,Stealth2015,Polarizability2015}.  What is
new here is the quantitative construction of a strongly screened elastic
current that overcomes both the xenon nuclear suppression and the falling
halo integral at the LZ candidate, and the conditional high-energy prediction
that comes with it.  We first ask, within a Gaussian-current
representation, how few independent terms an elastic form factor needs to
place the xenon recoil at $q_\star$ while respecting the standard LZ search;
the fewest we find is four, which matches the flavor content of an $SU(4)_D$
baryon, whose $U(1)_D$ charge form factor then acts as a momentum-space
band-pass filter---a statement about the representation, not about
microscopic constituent content.  We then confront a benchmark with LZ and extract predictions
independent of the fate of this event: a second recoil lobe capped by the
kinematic endpoint, and nodes whose recoil energies scale with the target as
$1/m_A$.

\textit{What the recoil requires.}---Let a dark vector $\AD$ of mass $m_{\AD}$
couple with strength $g_D$ ($\alpha_D=g_D^2/4\pi$) to the conserved current
of a composite ${\cal B}$ of mass $m_{\cal B}$, and mix kinetically with
hypercharge, $-(\epsilon/2\cos\theta_W)X_{\mu\nu}B^{\mu\nu}$.  After
diagonalization $\AD$ couples to the electromagnetic current with strength
$\simeq\epsilon e$, so coherent scattering is proton weighted.  With
$\sigma_p^{(0)}=16\pi\alpha\alpha_D\epsilon^2\mu_p^2/m_{\AD}^4$,
\begin{equation}
\frac{d\sigma_A}{dE_R}=\frac{m_A\sigma_p^{(0)}}{2\mu_p^2v^2}
Z^2F_A^2(q)\,|\fd(q)|^2\left(\frac{m_{\AD}^2}{m_{\AD}^2+q^2}\right)^2 ,
\label{eq:dsigma}
\end{equation}
where $F_A$ is the nuclear charge form factor in the Helm
parametrization~\cite{LewinSmith1996} (standard parameters, $s=0.9$~fm,
which includes the proton size), and $\fd$ is the dark charge form factor of
${\cal B}$, defined in Eq.~(\ref{eq:FD}) below, in which all model
dependence resides.  We use natural xenon, the standard halo~\cite{Baxter2021}
($\rho_\chi=0.3\gev\,{\rm cm}^{-3}$, $v_0=238$, $v_{\rm esc}=544$~km/s,
annually averaged $v_E=250+15\cos\phi$~km/s) and the LZ
efficiency~\cite{LZ2026}, parametrized by error-function turn-on and
roll-off crossing $50\%$ at $5.4$ and $269.9\kev$ (End Matter); accepted
spectra are integrated to $300\kev$, where the acceptance is $10^{-3}$.  At
$q\sim0.2$--$0.3\gev$ xenon is in its diffraction regime, with Helm minima at
$94$, $278$, and $555\kev$: $F_A^2$ falls from $1.5\times10^{-3}$ at
$200\kev$ to $1.6\times10^{-4}$ at $248\kev$, and the halo integral falls
with it, so a power-law amplitude $q^n$ produces a xenon spectrum peaking
below $220\kev$ for $n\le12$ and at $234\kev$ for $n=16$.

The xenon kernel is largest where the standard LZ search, below
$\approx55\kev$, reported no significant excess above its background model,
so $\fd$ must be strongly suppressed for
$q\lesssim0.12\gev$ and largest near $q_\star$: a band-pass filter.  To state
this without a model we define a target on the momenta LZ has probed,
$q<\qroi\equiv\sqrt{2m_{\rm Xe}\times300\kev}=0.271\gev$.  Maximizing the
rate near the candidate with no smoothness constraint is ill posed, so we use
the family $F_t=(q/\qroi)^n$, the leading small-$q$ behavior of a screened
composite with all its zeros at the origin.  Its xenon spectrum hardens with
$n$: the fraction within $\pm1\sigma$ of the candidate ($215.5$--$280.5\kev$)
rises from $0.9\%$ at $n=2$ to $38\%$ at $n=10$ and $49\%$ at $n=12$, the
mode from $29$ to $210$ and $220\kev$.  We take $n=10$ as the target,
bracketed by $n=8$ and $12$; the $\pm1\sigma$ window is a descriptive
measure of hardness, not an acceptance criterion.

\textit{Gaussian-current representation.}---A neutral composite built from
$N$ species of charged constituents has $\fd=\sum_iQ_iF_i$ with
$\sum_iQ_i=0$, where each conserved current
$J_i^\mu=\bar\Psi_i\gamma^\mu\Psi_i$ in $J_D^\mu=\sum_iQ_iJ_i^\mu$ has
$\langle{\cal B}(p')|J_i^\mu|{\cal B}(p)\rangle=(p+p')^\mu F_i(q^2)$ for a
scalar ${\cal B}$, $F_i(0)=1$ and $q^2\equiv|\mathbf q|^2$.  We represent
each $F_i$ by a single Gaussian,
\begin{equation}
\fd(q)=\sum_{i=1}^N Q_i\exp\!\left[-\frac{q^2r_i^2}{6}\right],\qquad
\sum_{i=1}^NQ_i=0,
\label{eq:FD}
\end{equation}
with $r_i^2=-6F_i'(0)$ the mean-square radius of the species-$i$ number
distribution, to be determined nonperturbatively in a fully specified theory.
Here $N$ counts independent radial terms, which need not equal the number of
constituent species (a non-Gaussian species contributes several terms, a dark
nucleus may hold several constituents of one species), so the bounds below
constrain the representation, not the constituent count.  A Descartes-type rule for exponential
sums~\cite{PolyaSzego} limits Eq.~(\ref{eq:FD}) to at most $N-1$ real zeros
in $q^2$, one of which neutrality spends at the origin, so $N$ terms afford
at most $N-2$ finite nodes.

Two terms cannot work.  With $Q_2=-Q_1$ and $x=q^2$,
$F_2=Q(e^{-ax}-e^{-bx})$, $0\le a<b$, is the charge-radius form factor of a
neutral composite, and $F_2(x)/x=Q\int_a^b e^{-tx}\,dt$ decreases with $x$,
so
\begin{equation}
|F_2(q)|/|F_2(q_\star)|\ \ge\ (q/q_\star)^{2}\qquad(q<q_\star)
\label{eq:N2bound}
\end{equation}
for every choice of radii: no two-term form factor suppresses the low-$q$
amplitude better than the pure $q^2$ law, which places $72\%$ of the accepted
xenon rate below $55\kev$, finite radii making it worse ($74$--$83\%$).  If the expected yield within
$\pm1\sigma$ of the candidate is fixed to one event, this shape implies
$\approx80$ accepted events below $55\kev$ under the adopted response,
whereas the LZ WIMP search saw no excess above
background~\cite{LZ2025WIMP} (a conditional statement, not a likelihood
analysis).  With
all zeros at the origin an $N$-term form factor behaves as $q^{2(N-1)}$ at
small $q$, so reproducing $q^{10}$ without nodes would need $N\ge6$; with
finite nodes, $N=3$ and $4$ must approximate it, and we fit them.

\begin{table}[t]
\caption{Best $N$-term fits of Eq.~(\ref{eq:FD}) to the $n=10$ target on
$q<\qroi$.  $D$ is the total-variation distance between accepted spectra
(the misplaced fraction of the rate); the spectral columns are the per cent
of the accepted rate below $55\kev$ and within $\pm1\sigma$ of the
candidate, and the mode in keV; $N_{\rm sb}$ is the number of \emph{true}
recoils in $350$--$700\kev$ per $0.99$ accepted events, with the range among
fits within $D_{\min}+0.02$.  $N=2$ is the bound (\ref{eq:N2bound}); the
fitted parameters are in Table~\ref{tab:pars}, and the benchmark of
Eq.~(\ref{eq:radii}) has $D=0.13$, $6.4\%$, $28\%$, $200\kev$,
$N_{\rm sb}=1.1$.}
\label{tab:N}
\begin{ruledtabular}
\begin{tabular}{cccccc}
$N$ & $D$ & $<55$ & $\pm1\sigma$ & mode & $N_{\rm sb}$ \\
\hline
target & 0 & 0.0 & 38 & 210 & --- \\
2 & 0.85 & $\ge72$ & $\le0.9$ & 29 & --- \\
3 & 0.39 & 5.6 & 11 & 170 & 0.1 (0.08--0.12) \\
4 & 0.11 & 5.5 & 29 & 201 & 1.3 (1.0--1.6) \\
\end{tabular}
\end{ruledtabular}
\end{table}

We fit radii and nodes (charges follow from neutrality and the node
conditions up to a scale) with $0.3\le r_i\le3$~fm and a cancellation floor
$|\fd(\qroi)|/\max_i|Q_i|\ge6.6\times10^{-4}$, minimizing by differential
evolution the total-variation distance
$D=\tfrac12\int_{5.4}^{300}dE_R\,|R_N-R_t|$ between the accepted spectra of
model and target, each normalized to unit integral (Table~\ref{tab:N};
settings, seeds, relaxed conditions, and a search over every node pattern in
the End Matter).  We call a
current \emph{hard} if its accepted spectrum lies within $D\le0.15$ of the
$n=10$ target.  No three-term form factor found is hard: its single finite
node settles at $0.098\gev$, the accepted spectrum peaks at $170\kev$ for
every seed and target considered, and $D\ge0.39$.  Four terms reach $D=0.11$
with nodes at $0.085$ and $0.166\gev$ (xenon recoils of $29$ and $113\kev$),
the accepted spectrum being controlled by the nodes together with the
Gaussian envelope.  Under this criterion and within this search domain, four
terms are the fewest that reproduce a hard target; more terms add freedom
and, as always, improve any fit~\footnote{Five terms reach $D=0.03$ with a
near-double node at $0.13\gev$; at a common tolerance the five-term family
contains the four-term one and widens the range of high-energy populations
in both directions (End Matter).  Nothing below depends on it.}.  The
criterion is one of template approximation, not of exclusion by the single
event: the three-term fit still places $11\%$ of its rate within $\pm1\sigma$
of the candidate, and a conditional single-event shape likelihood ratio
(signal only, fixed total expectation; End Matter) prefers the benchmark
below over it by a factor of only $2.4$ (over the two-term example by $30$).
A confining $SU(4)_D$ theory, whose baryon holds one constituent of each of
four flavors, is the simplest gauge realization of a four-term neutral
current that we know of.

\textit{A minimal realization.}---Consider $G_D=SU(4)_D\times U(1)_D$
with four vectorlike Dirac fermions
\begin{equation}
\Psi_i\sim({\bf 4},Q_i),\qquad
(Q_1,Q_2,Q_3,Q_4)=(3,-5,3,-1),
\label{eq:charges}
\end{equation}
a dark Higgs $\Phi$ of unit charge whose vacuum expectation value gives $\AD$
its mass, and the kinetic mixing above; the fermions are vectorlike, so all
anomalies cancel.  Eq.~(\ref{eq:charges}) is a compact integer
representative of the sign-alternating profiles the fits select; other
assignments, e.g.\ $(3,-4,2,-1)$, give the same nodes with different radii.  Flavor
breaking through $\bar\Psi_i\Psi_j\Phi_{ij}^{n_{ij}}/\Lambda_F^{n_{ij}-1}$,
with $n_{ij}=|Q_i-Q_j|$ and $\Phi_{ij}=\Phi$ or $\Phi^\dagger$ as gauge
invariance requires, lets heavier flavor baryons decay while dark baryon
number protects the lightest state~\cite{Appelquist2014,Stealth2015}; we work
to leading order in it.  Confinement produces the color-singlet baryon
\begin{equation}
{\cal B}_D=\epsilon_{abcd}\Psi_1^a\Psi_2^b\Psi_3^c\Psi_4^d ,
\label{eq:baryon}
\end{equation}
bosonic because the number of constituents is even and neutral,
$\sum_iQ_i=0$.  We restrict to the lowest $J^P=0^+$ channel, as in
``stealth'' dark matter~\cite{Appelquist2014,Stealth2015}, whose
vector-current matrix element contains only the charge form factor; that it
is the lightest baryon is an assumption.  The gauge theory motivates four
flavor-current contributions and their neutrality; representing each by a
single Gaussian is an additional assumption, and the radii, and the
cancellation among them, are inputs of the benchmark that follows.

\begin{figure}[t]
\includegraphics[width=\columnwidth]{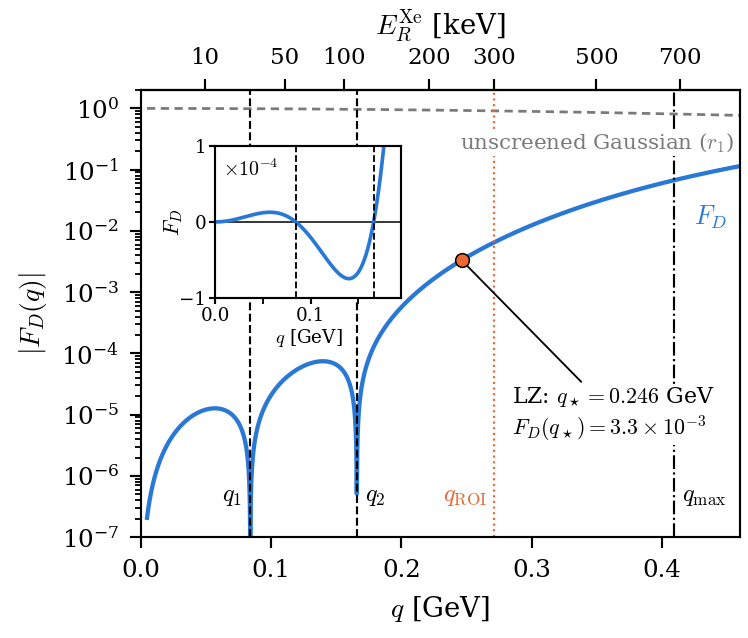}
\caption{Elastic $U(1)_D$ charge form factor of the benchmark,
Eqs.~(\ref{eq:charges}) and (\ref{eq:radii}); the top axis gives the xenon
recoil energy $q^2/2m_{\rm Xe}$.  Dashed lines mark the nodes $q_1=0.084$ and
$q_2=0.166\gev$, the dotted line the LZ acceptance edge $\qroi$, and the
dash-dotted line the kinematic endpoint $q_{\max}\simeq0.41\gev$ for
$m_{\cal B}=200\gev$; the LZ candidate probes $q_\star\simeq246\mev$, where
$\fd=3.3\times10^{-3}$.  Inset: $\fd$ on a linear scale over the screened band,
where $|\fd|<10^{-4}$ and the sign alternates across the nodes.}
\label{fig:FD}
\end{figure}

The benchmark radii,
\begin{equation}
r_i=(0.539,\ 0.805101,\ 1.271212,\ 1.574964)\ {\rm fm},
\label{eq:radii}
\end{equation}
were obtained by fixing $r_1$ and imposing nodes at $0.0842$ and
$0.1659\gev$ (those of the four-term fit to within $1\%$) and a screened
moment $\sum_iQ_ir_i^2=-1.95\times10^{-3}\,{\rm fm}^2$ (quoted to the
precision the cancellation requires; Fig.~\ref{fig:FD}); the nodes are thus
imposed by construction, not predicted by $SU(4)_D$.  The benchmark's
distance from the target is
$D=0.13$.  The cancellation is severe: the benchmark holds
$|\fd|\le1.3\times10^{-5}$ for $q<q_1$ and $\le7.5\times10^{-5}$ between the
nodes, against individual terms $|Q_iF_i|$ of order unity and
$\fd(q_\star)=3.3\times10^{-3}$---a cancellation to a part in $10^5$ across
$0<q\lesssim0.17\gev$, required because xenon's own form factor at $q_\star$
is only $\sim10^{-4}$.  The radii are load-bearing: under random fractional perturbations of
$10^{-4}$ ($10^{-3}$) the median accepted fraction below $55\kev$ rises to
$15\%$ ($73\%$), and exact leading-order screening alone,
$\sum_iQ_i\langle r_i^2\rangle=0$ with the higher moments fixed, removes both
nodes (End Matter).  The Gaussian shape itself is not what carries the spectrum: exponential or
Yukawa constituent profiles refitted to the same nodes and moment preserve
it ($9$--$11\%$ below $55\kev$, $40$--$42\%$ above $200\kev$), and fitted
directly to the target four terms of either profile reach $D=0.13$--$0.15$
while three stay at $0.40$ (End Matter).  A completion would
need dynamics enforcing a \emph{form-factor sum rule}, $\fd(q)\approx0$ over
the screened band, whose controlled breaking generates the second node; we do
not construct one, and establish the phenomenology conditional on this
screening.  The radii are of order a fermi while the baryon mass is hundreds of GeV;
since a heavy charged constituent would leave $\fd$ unscreened at high $q$,
the mass can be carried by $U(1)_D$-neutral heavy constituents or by binding
into a dark nucleus built from Eq.~(\ref{eq:baryon}), possibilities not
computed here that change the scattering state: throughout, the $200\gev$
object is an effective screened composite for which $SU(4)_D$ supplies a
possible current structure.

\textit{Recoil spectrum at LZ.}---For $m_{\cal B}=200\gev$, $m_{\AD}=0.5\gev$,
normalizing the accepted spectrum to $0.99$ events in $5.4<E_R<300\kev$---an
illustrative conditional spectrum, not a likelihood claim---fixes
\begin{equation}
\epsilon^2\alpha_D\simeq4.9\times10^{-12},
\label{eq:portalcomb}
\end{equation}
e.g.\ $\alpha_D=10^{-3}$, $\epsilon\simeq7.0\times10^{-5}$; the effective
strength at the LZ momentum,
$\sigma_p^{(0)}|\fd(q_\star)|^2[m_{\AD}^2/(m_{\AD}^2+q_\star^2)]^2
\simeq6.9\times10^{-44}\cm^2$, is far below
$\sigma_p^{(0)}\simeq9.7\times10^{-39}\cm^2$.  Of the accepted signal, $46\%$ lies above $200\kev$, $6\%$ below $55\kev$,
and $28\%$ within $\pm1\sigma$ of the candidate, with the mode near
$200\kev$, $1.5\sigma$ below the candidate as a consequence of the xenon
minimum at $278\kev$.  A Gaussian energy response of $23\sqrt{E_R/248\kev}\kev$ moves these
fractions by at most $2$ points and merges the $113\kev$ node with the
$94\kev$ Helm minimum into one trough (a $36\%$ deficit relative to a
node-free spectrum, separately resolved only for $\sigma\lesssim4\kev$); a
$\pm9\%$ energy-scale shift moves the fraction above $200\kev$ by
$\pm15$--$20$ points (End Matter).  LZ's selections on scintillation and
ionization observables are not modeled.

\begin{figure}[t]
\includegraphics[width=\columnwidth]{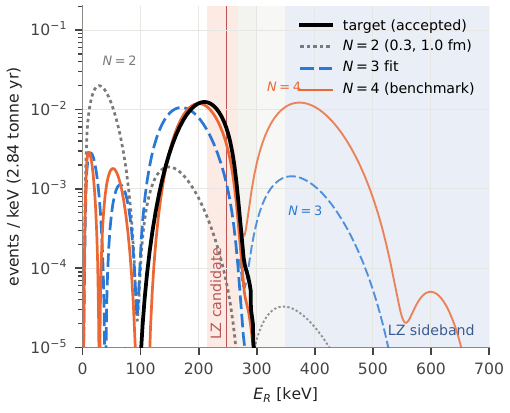}
\caption{Accepted xenon spectra (thick, $E_R<300\kev$) of the $n=10$ target,
a two-term example, the best three-term fit, and the four-term benchmark,
and their true-recoil spectra beyond the region of interest (thin),
normalized to $0.99$ accepted events.  The two-term example,
$(r_1,r_2)=(0.3,1.0)$~fm, places $74\%$ of the rate below $55\kev$ (by
Eq.~(\ref{eq:N2bound}) no two-term form factor places less than $72\%$
there); the three-term fit peaks at $170\kev$; the benchmark follows the
target ($D=0.13$) and places $1.1$ true recoils in the empty
$350$--$700\kev$ sideband.}
\label{fig:N}
\end{figure}

The true-recoil spectrum does not end with the region of interest: beyond
$q_\star$ the cancellation unwinds, $|\fd|$ grows from $3.3\times10^{-3}$ to
$0.07$ at the endpoint, and a second lobe at $300$--$500\kev$ as tall as the
accepted peak appears (Fig.~\ref{fig:N}), limited only by the kinematic
endpoint $E_R^{\max}=2\mu_{{\cal B}A}^2(v_{\rm esc}+v_E)^2/m_A\simeq0.69$~MeV
(for the maximal laboratory speed $809$~km/s).
The lobe is a property of the screened examples, not a theorem (End Matter).
The benchmark places $1.5$ true recoils beyond $269.9\kev$, growing from
$0.5$ at $m_{\cal B}=150\gev$ to $7.5$ at $1\tev$, so for a given form factor
this population is the observable that would bound the baryon mass.  LZ
reports an empty higher-energy sideband, $800<S1_c<1700$~phd (roughly
$350$--$700\kev$)~\cite{LZ2026,Rodd2026}, whose nuclear-recoil acceptance is
not published; a detected yield would require it, the energy reconstruction,
and a normalization profiled jointly with the region of interest, so the
true-recoil integrals quoted here are inputs to such an analysis, not its
result.  The benchmark places $1.1$ true recoils there ($3.1$ for $350\gev$);
as an illustration only, with unit acceptance and no background an empty
sideband would have Poisson probability $33\%$ for the benchmark and below
$5\%$ for $m_{\cal B}\gtrsim350\gev$.

The large $\sigma_p^{(0)}$ cannot be compared with pointlike
spin-independent limits, since $\fd(0)=0$.  With $\alpha_D=10^{-3}$ the
$\AD$ decays promptly ($c\tau\approx10~\mu$m), so the relevant bounds are the
prompt BaBar and LHCb searches~\cite{BaBar2014,LHCbDarkPhoton2018,
DarkPhotonReview}, well above this $\epsilon$ at $0.5\gev$.  The polarizability operator of stealth dark
matter~\cite{Polarizability2015} is not controlled by the screening, but the
massive mediator confines the field the nucleus sources to within
$1/m_{\AD}\simeq0.4$~fm of its surface: for a dark polarizability
$\alpha_E=\alpha_D(1~{\rm fm})^3$ the two-vector amplitude is at most
$2\times10^{-5}$ of the one-vector one across the stop band (End Matter).  The relic abundance is not fixed by the spacelike form factor;
an asymmetric population protected by dark baryon number~\cite{ZurekADM2014}
is the natural realization.

\textit{Target dependence.}---The transferable predictions are the
momentum-space nodes $q_{1,2}$: four terms admit at most two finite nodes,
the benchmark's are fixed by the fit to the xenon window, and the prediction
is that the same node momenta appear in every target, with recoil energies
$q_{1,2}^2/2m_A$ scaling as $1/m_A$ rather than with the reduced-mass
relation of inelastic scattering.  For natural xenon, argon, germanium, and
tungsten the nodes fall at ($29$, $113$), ($95$, $370$), ($52$, $203$), and
($21$, $80$)$\kev$, and $q_\star$ at $247$, $813$, $447$, and $177\kev$.
Tungsten is particularly favorable: both nodes and $q_\star$ lie below
$180\kev$, away from its own Helm minima, its halo integral at $q_\star$
exceeds xenon's, and both nodes survive a $5\%$ resolution, though
resolving them statistically needs far larger exposures (End Matter,
Fig.~\ref{fig:XeW});
germanium reaches $q_\star$ only from the halo tail; argon cannot reach
$q_\star$ with halo dark matter ($v_{\min}=1175$~km/s), so its test is the
first node at $95\kev$ ($v_{\min}\simeq400$~km/s).  Isotope mixtures, detector response, and each target's own nuclear
diffraction must be modeled; what neither a node-free form factor nor an
inelastic splitting produces is an additional structure common to all
targets in momentum space.

\textit{Conclusion.}---Within a Gaussian-current representation and an
explicit approximation criterion, a hard elastic recoil spectrum at LZ is
reproduced by no fewer than four independent terms, and the four-term
benchmark fixes a definite pair of momentum-space nodes; a confining
$SU(4)_D\times U(1)_D$ theory with a neutral four-quark baryon provides such
a current and motivates, without deriving, the benchmark.  Three consequences
follow.  First, the node momenta are common to all targets while the recoil
energies scale as $1/m_A$, so the same current predicts nodes at $95$ and
$370\kev$ in argon and at $52$ and $203\kev$ in germanium, a common
momentum-space structure that neither a node-free form factor nor an
inelastic splitting produces once each target's nuclear response is
modeled.  Second, the screened form
factor produces a second recoil lobe beyond the region of interest, capped by
the kinematic endpoint; we strongly encourage a joint signal-region and
sideband analysis of that lobe, which would bound the baryon mass.  Third,
the low-recoil stop band suppresses the rate below $55\kev$ to a few percent
of the total, so the absence of a conventional WIMP-like population at low
energies is expected rather than in tension.  The screening remains the mechanism's principal theoretical cost, to be
derived from additional dark flavor dynamics.
Neither the representation bounds nor the template comparisons infer
microscopic constituent content from a single event; the appropriate reading
of a $2.6\sigma$ candidate is a falsifiable proof of principle, and a
population of high-recoil events across targets would be decisive.

\begin{acknowledgments}
This work was supported by National Research Foundation of Korea (NRF) grants
funded by the Korean government (MSIT), Nos.~RS-2024-00340153 and
RS-2026-25607498.
\end{acknowledgments}

\bibliographystyle{apsrev4-2}
\bibliography{lz_su4_u1_refs}

% End Matter: keep two-column (no \onecolumngrid); PRL allows up to two pages here.
\section*{End Matter}

\textit{Appendix A: Fit procedure, coverage, and fitted parameters.}---For
$N$ terms the free parameters are the $N$ radii and $N-2$ node momenta; the
charges are the null vector of the $(N-1)\times N$ linear system of
neutrality and node conditions, normalized to $\max_i|Q_i|=1$.  Bounds are
$0.3\le r_i\le3$~fm and $0.02\le q_j\le0.45\gev$; the cancellation floor
$|\fd(\qroi)|/\max_i|Q_i|\ge6.6\times10^{-4}$ (the benchmark's value at
$q_\star$) excludes coincident-radius limits in which Eq.~(\ref{eq:FD})
mimics a power law with vanishing amplitude.  $D$ is minimized with
{\tt scipy.optimize.differential\_evolution} (population $15$ per dimension,
$\le300$ generations, tolerance $10^{-10}$, local polish) from seeds
$0,1,2$, which agree in $D$ to three decimals; the ranges in
Table~\ref{tab:N} extremize $N_{\rm sb}$ subject to $D\le D_{\min}+0.02$.  This parametrization imposes the maximal number of
finite nodes.  To cover neutral currents with fewer nodes we also imposed
$k=0,\ldots,N-3$ nodes with the charge vector on the unit sphere of the
remaining null space, and separately optimized $N-1$ free charge ratios with
neutrality alone, with the same bounds and floor.  Three-term currents reach
$D=0.394$ with one node in every parametrization (no node-free current does
better); four-term currents with $k=0,1$ converge to two-node solutions at
$D=0.147$ and $0.119$, above the $k=2$ value $0.113$; five-term currents
with $k=0,1$ reach $0.127$ and $0.133$ against $0.028$ with three nodes; the
free-charge search reproduces the three-term optimum and, ill conditioned
for a part-in-$10^5$ cancellation, finds two-node four-term solutions only
occasionally (best $D=0.40$, $0.15$, $0.16$ for the Gaussian, exponential,
and Yukawa profiles).  The minimality statement thus holds over the full
neutral $N$-term families within the stated bounds.  Lowering the floor to $10^{-5}$ gives $D=0.39$ and $0.097$
($N_{\rm sb}=1.7$) for three and four terms; a radius bound of $0.1$~fm
changes $D$ by $\le0.003$.  Five terms reach $D=0.027$ (nodes $0.083$,
$0.131$, $0.136\gev$; $N_{\rm sb}=3.8$, range $1.8$--$4.8$); at a common
tolerance $D\le0.13$ their range is $0.5$--$7.3$ against $1.0$--$1.5$ for four
terms, and for either $N$ the minimum $N_{\rm sb}$ rises as the tolerance is
tightened.  Table~\ref{tab:pars} gives the fits to the precision that
recovers them exactly.

\begin{table*}[t]
\caption{Best fits to the $n=10$ target at full precision: radii, node
momenta, charges (from neutrality and the nodes, relative to the smallest in
magnitude), $D$, and $N_{\rm sb}$; the printed radii and nodes reproduce $D$
to $10^{-5}$.  The benchmark is listed for comparison.}
\label{tab:pars}
\begin{ruledtabular}
\begin{tabular}{llllcc}
$N$ & $r_i$ [fm] & nodes [GeV] & $Q_i$ & $D$ & $N_{\rm sb}$ \\
\hline
3 & 0.300000, 0.424651, 0.602132 & 0.0979763 & $2.006203,\,-3.006203,\,1$ & 0.39414 & 0.113 \\
4 & 0.300001, 0.622341, 1.105829, 1.239918 & 0.0846797, 0.1656064 & $-1,\,1.731906,\,-1.731906,\,1$ & 0.11344 & 1.351 \\
bench. & 0.539, 0.805101, 1.271212, 1.574964 & 0.084200, 0.165898 & $3,\,-5,\,3,\,-1$ & 0.127 & 1.108 \\
\end{tabular}
\end{ruledtabular}
\end{table*}

\textit{Appendix B: Robustness and profile dependence.}---The random rows
quoted in the text use $r_i\to r_i(1+w\xi_i)$, $\xi_i$ standard normal from
{\tt numpy.random.default\_rng(1)} (four per draw in the order of
Eq.~(\ref{eq:radii}), $40$ draws per width, widths $10^{-4}$, $10^{-3}$,
$3\times10^{-3}$, $10^{-2}$ in successive blocks), charges fixed, each
spectrum renormalized: median fractions below $55\kev$ of $15$, $73$, $76\%$
(ranges $6$--$60$, $9$--$82$, $38$--$89\%$) at $10^{-4}$, $10^{-3}$,
$10^{-2}$, with $\epsilon^2\alpha_D$ spreading by factors of $3$, $300$,
$2000$.  The second lobe is not a theorem: a common envelope
$r_i^2\to r_i^2+2$~fm$^2$ preserves nodes and moment and lowers the
$350$--$700\kev$ population from $1.1$ to $0.4$ true recoils, at the price of
$38\%$ above $200\kev$ and $12\%$ below $55\kev$.  Moment tests: shifting $r_4$ alone by $-3.925\times10^{-4}$ sets
$M_2=\sum_iQ_ir_i^2$ to zero but also changes $M_4$ by $5.8\%$ ($47\%$ below
$55\kev$); $r_{2,3,4}=(0.804554,1.270722,1.574556)$~fm sets $M_2=0$ with
$M_4$, $M_6$ fixed and removes both nodes ($|\fd|\le2.5\times10^{-4}$ below
$0.17\gev$, $54\%$ below $55\kev$).
Profiles: $F_i=f(q^2r_i^2/6)$ with $f=e^{-x}$ (Gaussian), $(1+x/2)^{-2}$
(exponential density), and $(1+x)^{-1}$ (Yukawa density).  Refitting $r_{2,3,4}$ to the benchmark nodes and moment gives
$(0.735412,1.107386,1.359512)$ and $(0.700415,1.021174,1.244585)$~fm, fractions below
$55$/above $200\kev$ of $9/42\%$ and $11/40\%$, $D=0.16$ and $0.19$
(qualitatively hard, just outside the criterion because the Gaussian
benchmark's nodes and moment are kept), true recoils beyond $269.9\kev$ of
$1.04$ and $0.82$, and $\epsilon^2\alpha_D$ larger by $1.5$ and $2.0$.
Fitted directly to the target with their own radii and nodes, four-term
exponential and Yukawa currents reach $D=0.135$ and $0.151$ (nodes $0.084$,
$0.165$ and $0.083$, $0.164\gev$; $N_{\rm sb}=1.0$ and $0.8$) while
three-term currents of either profile stay at $D=0.40$: the criterion is
met, at its edge for the Yukawa profile, by four terms in all three profiles
and by three in none.

\begin{figure}[t]
\includegraphics[width=\columnwidth]{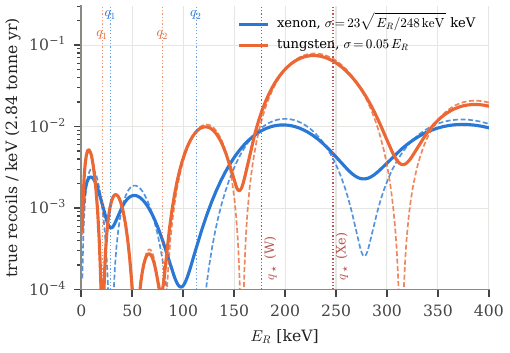}
\caption{True-recoil spectra of the benchmark in natural xenon and tungsten
at the same coupling, per $2.84$ tonne-years of each element's mass, before
(thin dashed) and after (thick) a Gaussian energy response of
$23\sqrt{E_R/248\kev}\kev$ for xenon and $5\%$ for tungsten.  Dotted lines
mark the common node momenta, $q_1$ and $q_2$, at $29$ and $113\kev$ in xenon
and $21$ and $80\kev$ in tungsten, and the candidate momentum $q_\star$ in
each target; the remaining dips are the targets' own Helm minima.}
\label{fig:XeW}
\end{figure}

\textit{Appendix C: Detector response.}---The LZ nuclear-recoil efficiency
of Ref.~\cite{LZ2026} is parametrized as
$\varepsilon(E_R)=0.955\,\Phi[(E_R-5.3\kev)/1.85\kev]\,
\{1-\Phi[(E_R-270.5\kev)/9.7\kev]\}$ with $\Phi$ the normal distribution
function, fitted to the published ``$+$ROI'' curve, which it reproduces to
$0.02$ (absolute) above $10\kev$; the roll-off coincides with the xenon
form-factor minimum at $278\kev$, so a linear roll-off from $250$ to
$300\kev$ changes every quoted fraction by less than $0.5$ points, and a
symmetrized-Fermi charge profile~\cite{Piekarewicz2016} in place of Helm
shifts the fractions by $2$--$4$ points.  We model the
reconstructed energy by a Gaussian of width $\sigma(E_R)=s_0\sqrt{E_R/248\kev}$
applied to the accepted spectrum, with $s_0=23\kev$, the candidate's
statistical uncertainty, as a proxy for the nuclear-recoil energy
resolution; the acceptance is applied in true energy and the smearing only
redistributes the shape, so no acceptance is counted twice, but the LZ
selections on $S1_c$ and $S2_c$ and the NEST response are not modeled.
Table~\ref{tab:resp} gives the reconstructed-energy fractions of the
benchmark and of the three-term fit under this response, at half and twice
the width, and under $\pm9\%$ energy-scale shifts, with the depth of the
trough in $78$--$148\kev$ (minimum over maximum) and the conditional
single-event shape likelihood ratio: with signal only and the total
expectation fixed, the likelihood of one event at $248\kev$ is proportional
to the smeared accepted density there, which relative to the benchmark is
$1.36$ for the target, $0.41$ for the three-term fit, and $0.03$ for the
two-term example under the nominal response ($0.35$--$0.57$ and
$0.02$--$0.05$ across the variations)---neither a background-inclusive
discovery test nor a comparison that accounts for parameter fitting.  The $78$--$148\kev$ interval contains both the xenon
Helm minimum at $94\kev$ and the dark node at $113\kev$ ($111.5\kev$ after
isotope averaging); under the nominal response they merge into a single
trough with its minimum at $98\kev$, so the tabulated depth measures the
combined nuclear and dark feature.  Against a node-free control (the $n=10$ template, same normalization and
response) the ratio of the smeared benchmark to the control falls to $0.64$
of its $150\kev$ value at $106\kev$ ($0.18$ at half the width, $0.05$ at
$\sigma(113\kev)=3.4\kev$), and the two minima appear separately only for
$s_0\lesssim5\kev$, i.e.\ $\sigma\lesssim4\kev$ at $113\kev$; the
transferable prediction is the zero of $\fd(q)$, and the minimum of a
smeared spectrum need not sit at $q_i^2/2m_A$.  Figure~\ref{fig:XeW} compares the
benchmark's true-recoil spectra in xenon and tungsten at the same coupling,
per $2.84$ tonne-years of each element's mass (not of a compound target): in
tungsten it gives $8.3$ true recoils, $7.7$ of them above $q_\star$ and only
$0.6$ below $177\kev$, where both nodes lie.  With an illustrative $5\%$
resolution the nodes survive as features, away from tungsten's own Helm
minima near $155$ and $310\kev$, but identifying them statistically would
require exposures of order $10^2$ tonne-years of tungsten.

\begin{table}[t]
\caption{Reconstructed-energy fractions (per cent below $55\kev$ / in
$200$--$300\kev$ / mode in keV) of the benchmark and of the three-term fit
under the response model of the text (nominal
$\sigma=23\sqrt{E_R/248\kev}\kev$), the depth of the benchmark's trough in
$78$--$148\kev$ (a combined nuclear and dark feature), and the conditional
single-event shape likelihood ratio of the three-term fit to the benchmark
at $248\kev$.}
\label{tab:resp}
\begin{ruledtabular}
\setlength{\tabcolsep}{3pt}
\begin{tabular}{lcccc}
response & benchmark & trough & three-term & $L_3/L_4$ \\
\hline
none & 6.4/45.8/200 & 0.001 & 5.6/21.4/170 & 0.32 \\
nominal $\sigma$ & 6.0/45.1/198 & 0.03 & 5.6/23.6/170 & 0.41 \\
$\sigma\times0.5$ & 6.3/45.5/200 & 0.003 & 5.6/22.0/170 & 0.35 \\
$\sigma\times2$ & 5.7/43.7/196 & 0.18 & 5.6/27.5/168 & 0.57 \\
scale $+9\%$ & 5.5/60.9/216 & 0.06 & 5.4/36.9/184 & 0.50 \\
scale $-9\%$ & 6.6/26.3/180 & 0.02 & 5.9/11.8/154 & 0.35 \\
\end{tabular}
\end{ruledtabular}
\end{table}

\textit{Appendix D: Polarizability estimate.}---With
${\cal L}\supset C_X{\cal B}_D^\dagger{\cal B}_DX_{\mu\nu}X^{\mu\nu}$ and
${\cal B}_D^\dagger{\cal B}_D\to n/2m_{\cal B}$ in the nonrelativistic
limit, the operator is a potential $V=-\tfrac12\alpha_E\mathbf E_X^2$ with
$\alpha_E=-2C_X/m_{\cal B}$, where $\mathbf E_X=-\nabla X^0$ and the nuclear
charge density sources the dark field through the kinetic mixing,
$X^0(\mathbf r)=\epsilon e\int d^3r'\rho_A(\mathbf r')
e^{-m_{\AD}|\mathbf r-\mathbf r'|}/4\pi|\mathbf r-\mathbf r'|$.  The Born
amplitudes of the two exchanges are, in magnitude,
$|\tilde V_1|=g_D\epsilon e\,|\tilde\rho_A(q)\fd(q)|/(q^2+m_{\AD}^2)$ and
$|\tilde V_2|=\tfrac12|\alpha_E|(\epsilon e)^2|S(q)|$ with
$S(q)=\int d^3r\,\mathbf E_1^2(\mathbf r)e^{i\mathbf q\cdot\mathbf r}$,
$\mathbf E_1$ the field per unit $\epsilon e$, and
$d\sigma/dE_R=m_A|\tilde V|^2/2\pi v^2$ in either case; the relative sign
(attractive for $\alpha_E>0$) and interference with the one-vector amplitude
are neglected, which is immaterial at the ratios found.  For a Fermi density
with the Helm parameters, $m_{\AD}R_A\simeq14$ confines the field to the
nuclear surface shell, so $S$ is of order $Z^2/(m_{\AD}^4R_A^4a)$ rather
than a volume integral, $S(0)=2.0\times10^{-3}\gev$.  With the benchmark
couplings and $\alpha_E=\alpha_D(1~{\rm fm})^3=0.13\gev^{-3}$,
$|\tilde V_2/\tilde V_1|$ is $2\times10^{-5}$ at $q=0.03\gev$,
$9\times10^{-6}$ at $0.05\gev$, and $1$--$2\times10^{-6}$ between the nodes
and at $q_\star$, exceeding unity only in the immediate vicinity of the exact
nodes.  Integrated with the halo and the acceptance, the two-vector rate
below $55\kev$ equals the one-vector rate there ($0.063$ events) for
$\alpha_E=1.7\times10^4\gev^{-3}$ and reaches one event in $2.84$
tonne-years for $\alpha_E=7\times10^4\gev^{-3}$, i.e.\
$|C_X|\lesssim7\times10^6\gev^{-2}$ at $m_{\cal B}=200\gev$; the two-vector
rate scales as $\alpha_E^2/\alpha_D^2$ at fixed $\epsilon^2\alpha_D$.  This
is a coherent classical-field estimate with a smooth nuclear charge density;
incoherent single-proton contributions and the validity of a local
polarizability operator at $q\sim m_{\AD}$ lie outside its scope, so it
supports the single-exchange treatment rather than excluding every
two-vector contribution.

\end{document}